# Automated Classification of Phonetic Segments in Child Speech Using Raw Ultrasound Imaging

Saja Al Ani[1], Joanne Cleland[2] and Ahmed Zoha[1]
[1]*James Watt School of Engineering, University of Glasgow, Glasgow G12 8QQ, U.K.*
[2]*School of Psychological Sciences and Health, University of Strathclyde, Glasgow, U.K.*

Abstract: Speech sound disorder (SSD) is defined as a persistent impairment in speech sound production leading to reduced speech intelligibility and hindered verbal communication. Early recognition and intervention of children with SSD and timely referral to speech and language therapists (SLTs) for treatment are crucial. Automated detection of speech impairment is regarded as an efficient method for examining and screening large populations. This study focuses on advancing the automatic diagnosis of SSD in early childhood by proposing a technical solution that integrates ultrasound tongue imaging (UTI) with deep-learning models. The introduced FusionNet model combines UTI data with the extracted texture features to classify UTI. The overarching aim is to elevate the accuracy and efficiency of UTI analysis, particularly for classifying speech sounds associated with SSD. This study compared the FusionNet approach with standard deep-learning methodologies, highlighting the excellent improvement results of the FusionNet model in UTI classification and the potential of multi-learning in improving UTI classification in speech therapy clinics.

## 1 INTRODUCTION

Speech sound disorder (SSD) is a common condition in early childhood, with a range of speaking difficulties affecting intelligibility (Shahin et al., 2019). Current approaches to the assessment and treatment of SSD rely on the perceptual skills of the treating clinicians, but this is known to be subject to difficulties with reliability and time-consuming. Technical solutions to this problem are required, particularly the automatic classification of images into specific speech sounds, for assessment and tracking progress in speech therapy. Utilising ultrasound tongue imaging (UTI) to visualise the movement and deformation of the tongue is currently a prominent technique in clinical phonetics that shows promise for the assessment and treatment of SSDs. This approach can image tongue motion at a relatively high frame rate of 60 Hz or higher, allowing for the observation of subtle and quick movements during speech production. Recent developments in the field of UTI have focused on feature selection and contour extraction (Xu et al., 2016). Despite these enhancements, accurate interpretation remains a challenge characterised by high-level speckle noise and information loss during dimension reduction (Zhu et al., 2018).

In response to these obstacles, researchers have made remarkable advances in deep learning using Convolutional Neural Networks (CNNs). CNN has become the method of choice for researchers investigating UTI processing, offering a solution that addresses tasks such as contour segmentation, feature selection, and tongue image classification, which are critical for enhancing the accuracy and efficiency of UTI analysis (Hueber et al., 2007) and, in turn, its clinical application in speech therapy clinics.

However, the efficiency of deep learning models requires a sufficient amount of labelled data, which is difficult to acquire in practice owing to the cost of labelling. Therefore, using multimodal learning with image and texture features can be beneficial, particularly in the healthcare sector, where the integration of medical images with another source of information can lead to more precise diagnoses and treatment recommendations. In this study, we explored UTI classification using a multi-learning data approach, including our proposed FusionNet model. This model combines UTI data with extracted texture features, utilising a combination of image and texture feature processing layers to enhance the analysis and classification of UTI. For the classification task, we employed various sets of deep-learning methodologies. This included CNN and DNN models, as well as pre-trained models, such as ResNet50 and Inception V3. Our inclusive approach involved the FusionNet model along with these established methods to demonstrate the efficacy of various techniques in UTI classification.

## 2 RELATED WORK

Inspired by advancements in deep learning, researchers have studied various supervised and unsupervised learning techniques to distinguish between ultrasound tongue motion. For instance, Hueber *et al*. and Cai *et al*. (Hueber et al., 2007), (Cai et al., 2011) recommended applying principal component analysis (PCA) and discrete cosine transform (DCT) to extract features in their classification tasks for silent speech synthesis and recognition. However, these feature representations may lose important details from the UTI during the overall dimension reduction procedure. Xu *et al*.(Xu et al., 2017) are an outliner because they use CNN to analyse tongue gesture classification from ultrasound data. However, this study only focused on two speakers, with generalisation to a third. Furthermore, an automatic approach for extracting the contour of the tongue from ultrasound data has been presented by Fabre *et al.* (Fabre et al., 2015). Using data from eight speakers for training and one held-out speaker for evaluation, the system was assessed in the speaker-independent mode. In each of these studies, a significant decrease in accuracy was observed when speaker-independent systems were used compared to speaker-dependent systems. You *et al.* discussed strategies for using several unlabelled UTI datasets to enhance the effectiveness of the UTI classification challenge (You et al., 2023). Using masking modelling, they investigated self-supervised learning. Their approach increased the classification accuracy in four different circumstances by an average of 13.33% compared with earlier competing algorithms.

In these studies, CNN models have been widely employed owing to their effectiveness and significant generalisation capacity. Achieving this robustness requires a sizable training dataset, which is rarely available when researchers employ their dataset. This study explored a multi-learning approach using two types of inputs. By combining the two types of inputs, the feature selection process can be significantly enhanced, leading to more promising results.

## 3 EXPERIMENTAL SETUP

### 3.1 Image Dataset

In this work, we utilised the Ultrax Typically Developing dataset (UXTD), which was obtained from the openly accessible UltraSuite repository (Eshky et al., 2018). This dataset was previously used in studies by (Ribeiro et al., 2019) and (Xiong et al., 2022). The dataset consists of a combination of phrases with words and phoneme speech data. For this study, only type A (semantically unrelated words) and type B (non-words) utterances were selected. Nine children's raw scan line data that represents the target utterances were extracted and transformed into 600x480x3 PNG images and four classes were determined to classify utterances:

1) bilabial and labiodental phones (e.g. /v/, /p/, /b/).
2) dental, alveolar and postalveolar phones (e.g. /th/, /d/, /t/, /z/, /sh/).
3) velar phones (e.g. /g/, /k/).
4) alveolar approximant /r/.

### 3.2 Texture Features Dataset

Texture analysis in ultrasound imaging plays a key role in the analysis of surface defect discovery (Xie, 2008) and image-based medical diagnosis (Castellano et al., 2004). In image processing, textural images refer to a specific pattern of distribution and dispersion of the intensity of the pixel illumination repeated sequentially throughout the image (Fekri-Ershad, 2019).

The process involves extracting features from an image based on its textural appearance and subsequently utilising these features for classification. In the current study, a Local Binary Patterns (LBP) operator was employed to demonstrate texture feature analysis. LBP is one of the textural image descriptors that can identify the local spatial structure and the local contrast of the image or part of it. It has become a broadly used texture descriptor due to its high classification accuracy in the implementation and extraction of proper features. This descriptor works by analysing each pixel with its neighbouring pixels by comparing them to a threshold value of its grayscale (Zhenhua Guo et al., 2010). The pixel serves as a centre of reference, and its grayscale level determines the classification of its neighbours as either 0 or 1. The centre pixel is then assigned a value which is a calculated sum of its binary neighbours:

$$\mathcal{L}_{P,R} = \sum_{p=0}^{P-1} sign(g_p - g_c)2^p \quad (1)$$

The above equation involves the grey levels of the centre and neighbouring pixels ($g_c$ and $g_p$), the total number of neighbourhood pixels (p), the radius (R), and binary values from thresholding. Figure 1 displays examples of the original UTI and the UTI after the LBP operator has been applied.

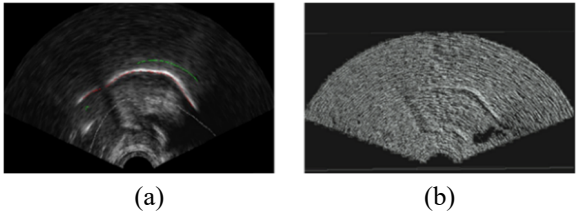

(a)            (b)

Figure 1: (a) original UTI, (b) UTI after performing LBP.

## 3.3 Classification Models

For the classification task, several deep learning methodologies have been developed, such as CNN and DNN models, which were adopted by Ribeiro *et al.* (Ribeiro et al., 2019). Furthermore, pre-trained models including ResNet50 and Inception V3 were involved, along with our proposed method, the FusionNet model.

A CNN is a specific type of neural network created to address image recognition problems. Convolutional layers are important because they can learn localized characteristics with a much smaller number of parameters. A collection of compact, locally receptive filters that convolve the entire input image are used in these layers. These filters are highly effective at detecting local characteristics because they analyse the relationships between pixels in smaller image areas. Pooling layers reduce spatial information by calculating averages over small regions in each feature map, thereby reducing the computational costs. Furthermore, fully connected are fed into the classification layer with fewer parameters, and therefore, less computing complexity.

The next model is the DNN, which is based on the perceptron model (Rosenblatt, 1958). Each node encounters several weighted inputs that are added to an activation function to produce the output value. This Perceptron can be integrated into a feedforward network, with the outputs of all nodes in one layer flowing into each node in the next, resulting in a completely connected network.

The ResNet50 architecture is a convolutional neural network with 50 layers deep (He et al., 2015). The main intention of using ResNet50 is the ease of optimisation and the fact that it has been trained on more than a million images from the ImagNet database (Krizhevsky et al., 2012), which makes it useful when we have limited data.

Furthermore, the Inception V3 is a convolutional neural network with 48 layers (Szegedy et al., 2015). It is known for its efficiency in capturing features at multiple scales and performance in image classification and computer vision tasks. The final layer of both pre-trained models was adapted to be compatible with the number of classes.

Finally, our proposed multi-learning method uses the FusionNet model. This model consists of two main parts: image processing layers for handling visual information, and texture descriptor processing layers for including additional texture-related features. The image layers utilise convolutional operations and max pooling for hierarchical feature extraction, whereas the texture layers comprise fully connected neural network segments. The outputs from these parts were concatenated and passed through fully connected layers with dropout regularization, leading to the final classification output. This architecture allows the model to leverage both visual and texture information to improve the performance in classification tasks.

All network architectures used in this study were optimised for 50 epochs using the stochastic gradient descent (SGD) optimiser at a learning rate of 0.001 and 32 mini-batches. After the training phase, the testing procedure begins by loading the test data. A dataset of randomly selected images and texture features is created to test the models. These data inputs were fed into the proposed models to predict the test dataset. Comparisons were made between the outcome values of each model testing phase.

## 3.4 Learning Scenarios

To compare our approach with other deep learning algorithms, we utilised data from nine speakers sourced from the UXTD dataset, a dataset previously employed by (Ribeiro et al., 2019) in their research. In our experiments, several scenarios were considered. First, it is speaker-dependent, where the training process is customised based on the unique samples of an individual speaker. Second, in the multi-speaker scenario, the system was trained using a dataset that included UTI samples from multiple speakers. The goal is to develop a model that can recognise and adapt to a variety of speakers. Third, speaker-independent systems aim to be more adaptable by being trained on a broader range of speaker samples; however, the main objective is to develop a model that can generalise well across speakers without being specifically tuned to any speaker's characteristics.

Three training and testing stages were performed. In step 1, only the UTI input using the DNN and CNN models was employed. In step 2, we further refined the CNN and DNN model architectures to enhance the model performance and utilise the pre-trained ResNet50 and Inception models. Finally, in Step 3, we investigated the performance of our proposed FusionNet model, which combined images and texture features as an input to train a deep learning model.

## 4 RESULTS

We trained five models at different stages using two different training input setups. For each stage, we present the results for every network structure to demonstrate how different architectures may vary the performance of the proposed deep learning model.

The results of step 1, UTI using CNN and DNN models, are shown in Figure 2, presenting an accuracy comparison with previously published data (Ribeiro et al., 2019). When comparing the model classifier, we observed that the CNN classifier outperformed all scenarios. Examining training scenarios, speaker-dependent systems demonstrate better performance at 74.30% accuracy compared to multi-speaker systems with 72.42% accuracy, which shows that the system adapted its learning to the unique attributes of a specific speaker. Speaker-independent systems underachieve, which explains the challenge involved in generalisation to unseen speakers.

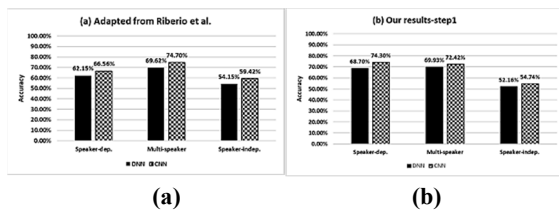

Figure 2: Accuracy scores for DNN and CNN models (a) previously published data[11], (b) our results.

After modifying the CNN and DNN models in Step 2, we reevaluated the classification. Furthermore, we conducted an assessment using ResNet50 and Inception V3 for all the speaker scenarios. Figure 3 shows the precision results for different scenarios. In this case, CNN and Inception V3 consistently demonstrated higher precision values across speaker-dependent, multi-speaker, and speaker-independent regions than DNN and ResNet50. DNN shows competitive precision in speaker-dependent and multi-speaker scenarios but experiences a significant drop in precision for speaker-independent scenarios.

ResNet50 performed well in speaker-independent scenarios, displaying higher precision values; however, CNN and Inception V3 maintained comparable performance in multi-speaker settings. Inception V3 outperformed it with consistently high precision across various scenarios. The precision results across different scenarios reveal notable distinctions among the evaluated models for UTI classification, especially in scenarios with diverse speaker characteristics.

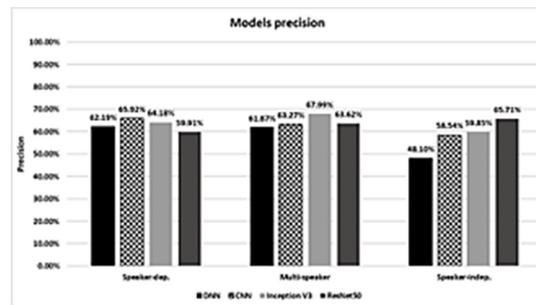

Figure 3: Models' precision performance.

Owing to the observed declines in precision for particular models and scenarios, in step 3, an alternate methodology was introduced and examined to improve the performance of the UTI classification. Figure 4 presents the precision performance results of the proposed FusionNet model across the speaker scenarios. In the speaker-dependent set, the model demonstrated a high precision of 91.88%. For the multispeaker scenario, the model demonstrated robust performance with a precision of 92.12%. In particular, in a challenging speaker-independent scenario, the model successfully achieved a precision of 82.32%.

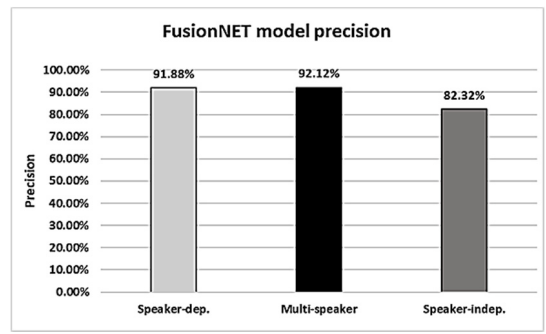

Figure 4: FusionNet model precision performance.

The behaviour of the training and testing losses and accuracy for speaker-independent are shown in Figure 5, where the losses decrease slowly, and the testing accuracy reaches almost its highest accuracy after 45 iterations and stabilises in further iterations with steady improvement.

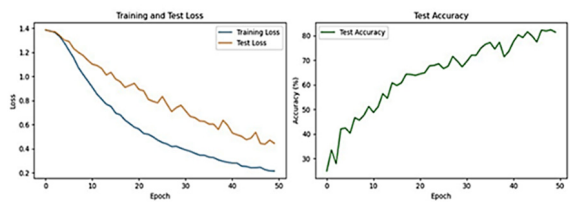

Figure 5: The accuracy and loss behaviours of the training and testing.

The high accuracy of the testing data is a sign of the success of the classifier. Confusion matrices were implemented to examine the performance of the FusionNet model further. In the confusion matrices, the row represents the actual utterances class, and the column represents the utterances class predicted by the model Figure 6 presents the confusion matrices where the model successfully achieved high accuracy in classifying the classes in the speaker-independent scenario. However, the most misclassified images in the dental-alveolar class scored relatively low accuracy compared to the other classes.

The implementation of the FusionNet model was shown to be an approach to the initial performance challenges, leading to a significant improvement in precision.

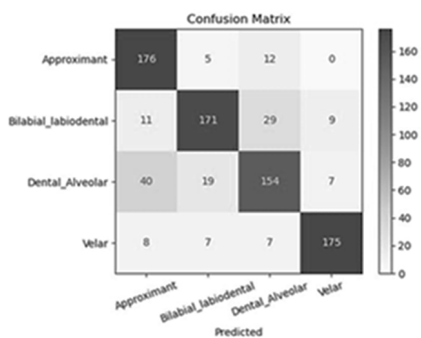

Figure 6: Confusion matrix for speaker-independent on the testing dataset.

## 5 CONCLUSION

In recent years, deep learning methodologies, particularly CNN, have been applied across diverse domains, including the diagnosis of speech disorders, phonetics studies, and segmentation of the tongue. The success of these techniques in speech fields has encouraged the idea of conducting this research by employing deep learning techniques for phonetics segment classification. In this study, image processing and deep learning algorithms have shown promising results in classifying UTIs from child speech. Accurate classification of UTI from child speech can be used for the automatic assessment of child speech. The performance of adapting different methodologies has been promising, although it degrades when evaluating previously unseen data, thereby emphasising the need for robust adaptability. An encouraging approach for improving the classification precision in all speaker scenarios was developed through the integration of multi-learning data. In particular, speaker-independent results showed excellent improvement, with a precision of 82.32%. To provide more clarity on the reported findings, future research should focus on two main aims. First, it seeks to expand the size of the dataset by including more samples. Second, it investigates which speaker scenarios or patterns contribute to classification errors.